\documentclass[10pt,a4paper]{article}
\usepackage[margin=0.74in]{geometry}
\usepackage[T1]{fontenc}
\usepackage[utf8]{inputenc}
\usepackage{lmodern,microtype,amsmath,amssymb,booktabs,array,graphicx,caption,subcaption}
\usepackage[colorlinks=true,urlcolor=blue,citecolor=blue,linkcolor=blue]{hyperref}
\usepackage{natbib}
\hypersetup{pdftitle={Early Assessment Submission and Subsequent Course Outcome in OULAD},pdfauthor={Devis Wawan Saputra}}
\title{Early Assessment Submission and Subsequent Course Outcome:\\ \large A Day-30 Landmark Observational Analysis of OULAD with Explicit Weighting Diagnostics}
\author{Devis Wawan Saputra\\\small Independent Researcher, Indonesia\\\small ORCID: \href{https://orcid.org/0000-0002-7133-9410}{0000-0002-7133-9410}}
\date{7 November 2026}
\begin{document}\maketitle
\begin{abstract}
We examined the association between submission of an assessment by day 30 and final course outcome in the Open University Learning Analytics Dataset (OULAD). To avoid classifying students who withdrew during the exposure window as comparable non-submitters, the main analysis was restricted to students enrolled at a day-30 landmark. In the CCC/2014J course presentation, 1,820 students remained after exclusions for missing covariates and unsupported exposure categories; 1,624 had submitted at least one non-banked assessment by day 30. The difference in favorable final outcome rates (Pass or Distinction) was 0.3743. A normalized inverse-probability-weighted contrast was 0.3861 (95\% full-refit bootstrap interval 0.3242--0.4447). The overlap-weighted contrast was 0.3261 (0.2667--0.3870), describing a different weighted population. The weights were unstable in the smaller comparison group, and measured covariates remained imbalanced under inverse-probability weighting. We independently reconstructed the cohort and estimates from the official UCI archive, examined eligibility and missing-data exclusions, and repeated descriptive analyses in three other course presentations. Those comparisons also showed positive associations, but their confidence intervals cannot account for unmeasured differences between the groups. The study identifies a reproducible association among selected learners, not the effect of requiring or encouraging earlier submissions.

\end{abstract}
\section{Introduction}
An assessment submission is easy to observe in administrative data. It is harder to interpret. Students who submit during the first month of a course may differ from those who do not in preparation, available time, access to assessment tasks, or progress toward course completion. Some of those differences may also predict the final outcome. Treating submission by day 30 as a fixed baseline characteristic can introduce further problems if students who withdraw before day 30 are included among the non-submitters.

We therefore compare students who remained enrolled at a common day-30 landmark. Our question is deliberately descriptive: within that selected population, how much do recorded rates of favorable final outcomes differ between students with and without a qualifying submission by day 30, after adjustment for available background variables? We examine the reliability of that comparison by reproducing it independently and inspecting weight concentration, covariate balance, exclusions, and other course presentations.

The analysis does not identify an effect of submission policy. An observed submission is neither a randomized assignment nor a sufficiently specified intervention. In particular, a course assessment may itself be part of the requirements for passing, making submission status and successful completion structurally related.

\section{Related Literature and Causal Structure}
OULAD links student characteristics, registration records, assessments, and course outcomes across Open University presentations \citep{kuzilek2017}. Landmark designs align the observation window and eligibility at a common time, reducing certain temporal classification problems but restricting the population under study \citep{hernan2016}. Inverse-probability weighting can adjust observed contrasts for recorded covariates, although model misspecification, weak overlap, and unmeasured factors still threaten interpretation \citep{austin2015}. Overlap weighting gives more influence to students with intermediate estimated probabilities of early submission and consequently targets a different population \citep{li2018}.

Let $E$ indicate a qualifying submission by day 30, $Y$ indicate Pass or Distinction, and $X$ denote recorded background variables. Motivation, work obligations, assessment opportunity, and prior preparation may influence both $E$ and $Y$ without being measured in $X$. Enrollment at day 30 can also depend on those factors. Conditioning on enrollment therefore creates a selected group; weighting its measured characteristics does not recover an unconfounded causal effect. This analysis follows relevant observational reporting principles from STROBE \citep{strobe2007}. Recent OULAD prediction studies examine early-warning calibration and fairness \citep{wu2026} and early academic-risk prediction from the first four weeks of activity and assessment \citep{wufourweeks2026}. These address predictive reliability rather than the effect of changing assessment submission behavior.

\section{Source Data and Cohort Construction}
\subsection{Provenance}
The analysis uses OULAD, UCI repository entry 349 and DOI \href{https://doi.org/10.24432/C5KK69}{10.24432/C5KK69}. The independent verification workflow downloaded the original UCI ZIP archive and verified SHA-256 \texttt{f2ed1902616c1fe8d2824d872c0b7d2d}\\\texttt{72be435bf0124d077044fe4be2c6d3e4}; tables used are \texttt{studentInfo.csv}, \path{studentRegistration.csv}, \texttt{assessments.csv}, and \path{studentAssessment.csv}. The original dataset description is \citep{kuzilek2017}. The dataset records historical 2013--2014 UK distance-learning modules; the analysis does not require post-landmark clickstream logs. Findings cannot automatically be transported to contemporary learning environments.

\subsection{Eligibility, exposure, outcome, and selection}
The evaluation landmark is presentation day 30. An included learner must be registered no later than day 30 and must not have unregistered by the end of day 30. For records whose final status is Withdrawn, an observed unregistration date after day 30 is required. We explicitly condition on survival in the course to day 30; estimates do not target all originally enrolled students.

Exposure $E=1$ requires at least one \emph{non-banked} submission, with the submission date no earlier than the recorded registration date and no later than day 30. Banked results transferred from earlier attempts are excluded. Outcome $Y=1$ is a final Pass or Distinction; $Y=0$ is Fail or a withdrawal occurring after the landmark. The binary unexposed category $E=0$ combines students who submit later and students who never submit a qualifying assessment: these are behaviorally different groups. That distinction cannot be recovered by interpreting $E=0$ as ``never submitted.'' A companion analysis script classifies these subgroups for \emph{descriptive} comparison, explicitly treating post-landmark submission as a future event that must not be used as a baseline confounder. Furthermore, submission is partly an assessment-progression process. Comparisons of submission status against successful completion risk capturing structural co-dependence, as well as learner behavior.

\subsection{Missingness and empirical support}
The independently reproduced source spans 32,593 module-registration records; 27,350 were landmark eligible across all presentations. The chosen CCC/2014J presentation contains 1,983 eligible learners before adjustment missingness. A further 133 cases were excluded owing to missing deprivation category, leaving 1,850 complete cases. The independently implemented categorical-support restriction removed 30 learners associated with categorical levels lacking empirical variation in the binary exposure, yielding 1,820 analytic cases. This outcome-blind restriction reduces unsupported comparisons but changes the target population and does not guarantee propensity overlap for every multivariate covariate pattern. The total restriction excludes 163 of 1,983 eligible learners (8.22\%). The independently audited unadjusted favorable-outcome contrast is 0.376763 among all 1,983 landmark-eligible learners, 0.376638 among the 1,850 complete cases, and 0.374296 among the final 1,820. These numerical similarities reduce concern that the measured \emph{unadjusted} contrast is solely an artifact of those two exclusion steps. They do not eliminate selection into the day-30 survivor cohort or explain the unmeasured causes of early submission. Missing-covariate exclusion affected 122 of 1,776 early submitters (6.87\%) and 11 of 207 non-early learners (5.31\%).

\begin{table}[ht]\centering\small\caption{Independently replicated flow for CCC/2014J at day 30.}\begin{tabular}{lr}\toprule
Selection step & Learners\\\midrule
Landmark eligible & 1,983\\
Missing adjustment covariate & $-133$\\
Complete case & 1,850\\
Empirically unsupported category & $-30$\\
Final analytic set & \textbf{1,820}\\
Early submitters / no early submission & 1,624 / 196\\\bottomrule
\end{tabular}\end{table}

\subsection{Independent exclusion audit and raw outcome sensitivity}
We separately tabulated submission status and final outcomes for the 133 learners missing a baseline covariate and the 30 excluded by the empirical-support rule. The audit used the checksum-verified UCI archive (\href{https://github.com/devissaputra/causal_learning_analytics/actions/runs/39688599327}{GitHub Actions run 39688599327}). These are unadjusted, retrospective descriptions, not imputed adjusted estimates.
\begin{table}[ht]\centering\small
\caption{New independent OULAD source-data exclusion audit (CCC/2014J, day 30). Rates are favorable final outcomes among the named observed exposure groups.}
\label{tab:exclusions}
\begin{tabular}{lrrrrr}\toprule
Set & $n$ & Early $n$ & Not early $n$ & Early rate & Not-early rate\\\midrule
All landmark eligible & 1,983 & 1,776 & 207 & 0.5507 & 0.1739\\
Missing covariates & 133 & 122 & 11 & 0.6967 & 0.3636\\
Support excluded & 30 & 30 & 0 & 0.6667 & n/a\\
Final analytic sample & 1,820 & 1,624 & 196 & 0.5376 & 0.1633\\\bottomrule
\end{tabular}\end{table}
The all-eligible \emph{raw} early-minus-not-early favorable-rate difference is $0.550676-0.173913=0.376763$; after the exclusion sequence it is $0.537562-0.163265=0.374296$. This small change is a direct calculation from observed final labels, not a sensitivity result for the \emph{adjusted} IPW or overlap estimands. Missing-covariate exclusions have different outcome rates from the final analytic sample, and the empirical-support step removes only early submitters; selection remains a material interpretive limitation. No claim of missing-at-random, zero selection bias, or preserved causal effect follows from the raw-difference comparison.

\section{Statistical Analysis}
\subsection{Adjustment and estimated propensity}
The numeric adjustment set includes studied credits, number of prior attempts, and registration timing. Categorical variables are age band, highest education, deprivation band, disability indicator, gender, and region. Numeric variables are standardized and categories are encoded using one-hot indicators. The exposure-propensity model is an $L_2$-penalized logistic regression $e(X)=P(E=1\mid X,L=1)$ fitted in the selected cohort. Assessment score, activity after day 30, and final outcome are excluded from adjustment. The available baseline records do not capture assignment due dates or individual assessment opportunities, pre-course academic ability, self-regulation, or motivation. The propensity model cannot balance these missing determinants; its covariate adjustment is neither a substitute for a randomized policy comparison nor a well-defined counterfactual intervention.

\subsection{Weighted contrasts and their target populations}
Let $w_i=E_i/e_i+(1-E_i)/(1-e_i)$. The normalized H\'ajek ATE-style weighted outcome contrast is
\begin{equation}
\widehat\Delta_{\mathrm{IPW}}=\frac{\sum_i E_iY_i/e_i}{\sum_i E_i/e_i}-\frac{\sum_i(1-E_i)Y_i/(1-e_i)}{\sum_i(1-E_i)/(1-e_i)}.
\end{equation}
This would target a standardized average causal effect only under consistency, positivity, correctly specified estimation, conditional exchangeability, appropriate selection assumptions, and well-defined interventions. Here it is reported descriptively as an ATE-\emph{style} weighted contrast.

Overlap weights are $w_i^{\mathrm{OW}}=E_i(1-e_i)+(1-E_i)e_i$. They emphasize subjects with intermediate estimated exposure probability. Their standardized difference targets the propensity-overlap population (often called an ATO-style target), not the same empirical average as IPW. No difference between the two estimates should automatically be described as evidence for effect heterogeneity, because the assumptions and weighting diagnostics also differ.

Covariate balance is summarized by standardized mean differences (SMD). Large SMD, narrow effective sample size (ESS), and extreme weights warn that weighted associations may be dominated by a relatively small set of observations. Kish ESS is $(\sum_i w_i)^2/\sum_i w_i^2$. The primary recorded uncertainty interval is obtained by 1,000 bootstrap replicates refitting the propensity model in each draw; a fixed-propensity bootstrap is retained only as a secondary diagnostic. Such intervals quantify sampling/model-fit variability under the frozen cohort and specifications, \emph{not} confounding uncertainty.

\section{Main Results}
\subsection{Outcome contrasts and uncertainty}
\begin{table}[ht]\centering\small\caption{Independently reproduced CCC/2014J contrasts. All are observational; last two intervals are full-refit percentile bootstrap intervals.}\label{tab:effects}
\begin{tabular}{lrrl}\toprule
Measure & Difference & 95\% interval & Population/interpretation\\\midrule
Raw early minus not-early & 0.3743 & not estimated & Unweighted final analytic set\\
Horvitz--Thompson & 0.3915 & not estimated & IPW, ATE-style\\
H\'ajek IPW & \textbf{0.3861} & [0.3242, 0.4447] & IPW, ATE-style\\
Overlap weights & \textbf{0.3261} & [0.2667, 0.3870] & ATO-style, different target\\\bottomrule
\end{tabular}\end{table}
\begin{figure}[ht]\centering\includegraphics[width=.79\linewidth]{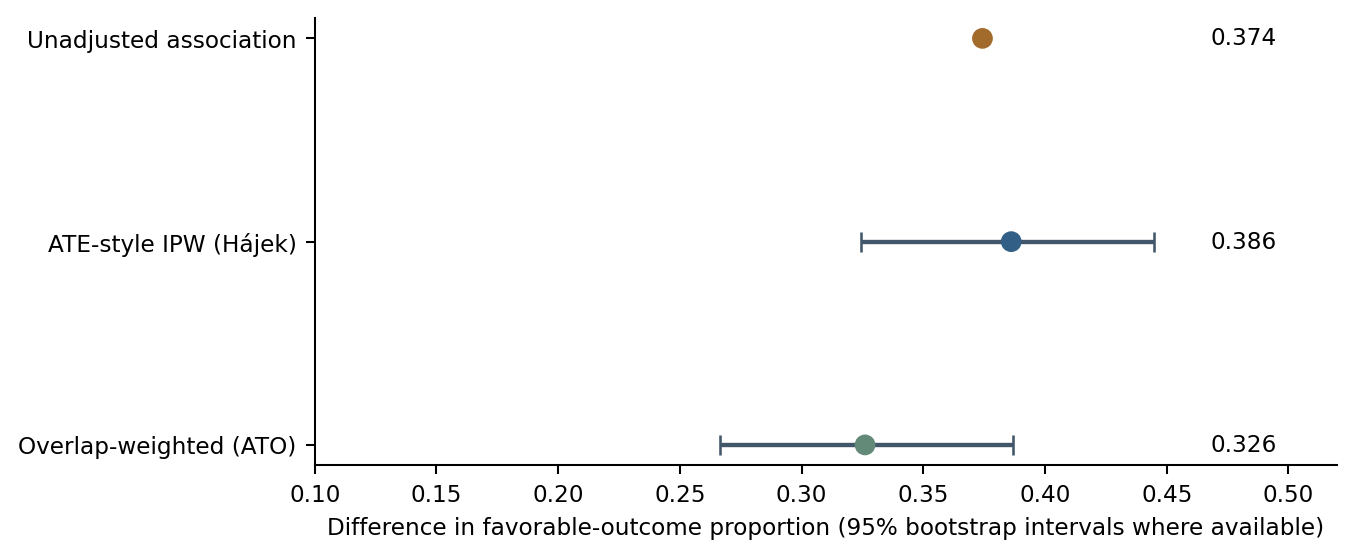}\caption{Archived numerical contrasts with bootstrap uncertainty, plotted from existing summary metrics. Raw and weighted estimates do not share the same estimand.}\end{figure}
Within the selected landmark cohort, earlier submission is associated with a higher recorded probability of a favorable final result. The adjusted contrast is numerically large, but it must not be read as an expected effect of a deadline, a reminder, a tutoring intervention, or compulsory assessment completion.

\subsection{Propensity overlap, effective sample size, and covariate balance}
The estimated propensities range from 0.3841 to 0.9915; the observed common-support interval is 0.4459--0.9741 and contains 97.03\% of cases. Under standard IPW the maximum weight is 38.64 and total ESS is about 509.52, despite 1,820 observed learners. The weighted effective sample size for the 196 unexposed learners is 133.81. These figures indicate substantial dependence on large weights in the sparse comparator group.

Several SMDs improve with IPW but not all satisfy a $|\mathrm{SMD}|<0.1$ descriptive threshold: for example female gender falls from absolute SMD 0.307 to 0.116, still above this threshold. Overlap weighting has maximum recorded absolute SMD 0.0181, total ESS 649.71, and unexposed-group ESS 192.32. Because the fitted propensity regression includes an $L_2$ penalty, exact finite-sample moment balance is not guaranteed even under the usual logistic overlap-weight formulation \citep{li2018}. This better measured balance is associated with a change in target population; it does not establish exchangeability or resolve unknown confounding.
\begin{figure}[ht]\centering\includegraphics[width=.79\linewidth]{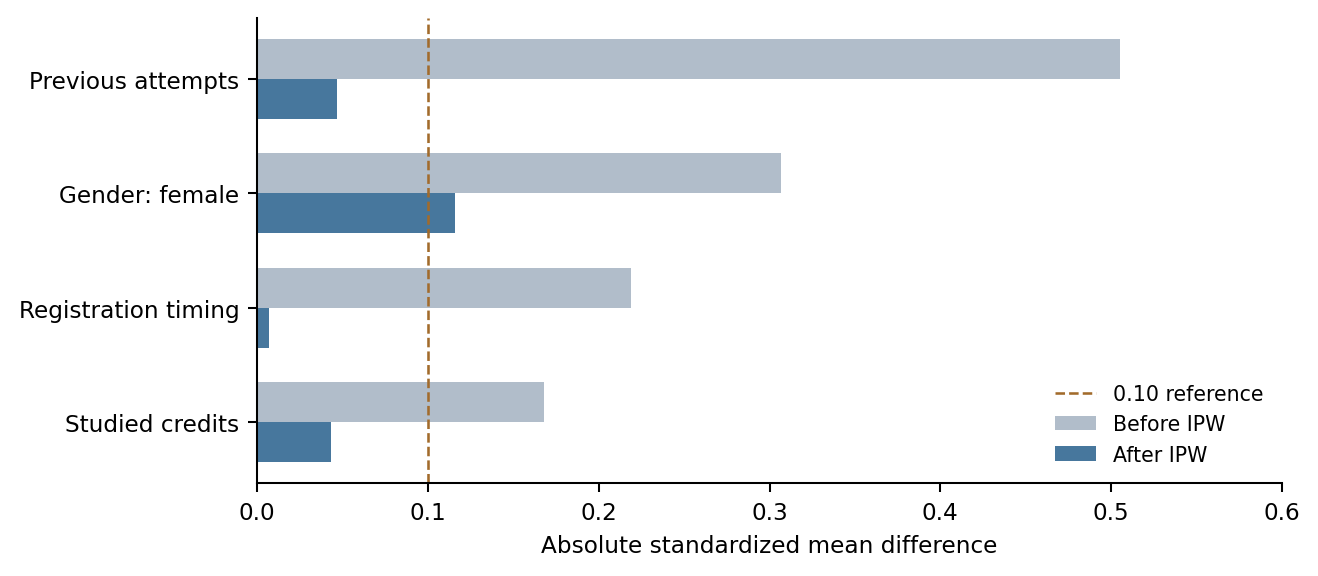}\caption{Selected absolute SMDs before and after primary IPW, reconstructed from existing empirical summary. Remaining imbalance is visible.}\end{figure}

\subsection{Recorded sensitivity to analytic choices}
The propensity-specification sensitivity analyses produced H\'ajek contrasts 0.3912 without region, 0.3801 when adding quadratic numeric terms, and 0.3849 under stronger $L_2$ regularization (primary: 0.3861). Clipping probabilities at 0.05 produces 0.3808, and clipping at 0.10 produces 0.3771, the latter changing weights for 1,101 cases. The empirical common-support restricted contrast is 0.3829, with scores not re-estimated after restriction. Such stability across selected modifications is descriptive and does not address confounding by motivation or preparation. Moreover, the overlap-weighted contrast (0.3261) applies to a different population and should not be marketed as a robustness confirmation of the same estimand.

\section{Illustrative Unmeasured-Confounding Scenario}
The bootstrap intervals describe variability from resampling students and refitting the propensity model; they do not quantify uncertainty about unmeasured confounding. To illustrate the scale of the problem, consider an additive bias term $B=\gamma\eta$, where $\gamma$ is a stipulated outcome-risk difference associated with an unmeasured binary factor and $\eta$ is its stipulated weighted prevalence difference between the submission groups. Both are hypothetical. Subtracting $B=0.20$ from the observed IPW contrast of 0.3861 would leave 0.1861; using $B=0.40$ would reverse its sign. The data do not establish whether either assumption is plausible. This calculation is an algebraic illustration, not a sensitivity bound estimated from OULAD, and it does not turn the descriptive comparison into a causal estimate.

\section{Independent Reproduction and Landmark Checks}
On 22 October 2026, we ran an independently written extraction, cohort-construction, propensity-estimation and outcome-analysis program on a fresh GitHub Actions runner. The program does not import the original project's causal analytics functions. It downloaded and checksum-verified the official UCI archive, extracted only the four required linked CSVs, calculated each file SHA-256, and independently rebuilt the day-30 CCC/2014J analytic cohort and primary contrasts. The run and aggregate JSON record are publicly auditable at
\url{https://github.com/devissaputra/causal_learning_analytics/actions/runs/39684326436}. The verification code is proposed in a separate, unmerged pull request \url{https://github.com/devissaputra/causal_learning_analytics/pull/1}. Raw student records are not bundled with this manuscript.

The 1,820 learner sample, split (1,624 vs. 196), unadjusted risk difference (0.374296), H\'ajek IPW contrast (0.386110), and overlap-weighted contrast (0.326066) reproduce the archived point estimates to the stated tolerances. The independent 1,000-resample full-refit percentile intervals are 0.324225--0.444675 and 0.266654--0.387026, respectively. The IPW effective sample size of 509.52 and maximum weight of 38.64 were also reproduced. Consequently, the previous uncertainty about whether the reported day-30 estimates were reproducible from raw OULAD records is resolved for these primary numerical quantities. This verification is not an independent causal identification exercise.

A pre-specified exploratory landmark check with the same data-processing logic yielded the following results:
\begin{table}[ht]\centering\small
\caption{Independently computed CCC/2014J landmark checks. Comparisons across days refer to different eligible learner populations. Day-45 point estimates do not have separately refitted bootstrap intervals.}
\begin{tabular}{lrrrrrr}\toprule
Landmark & Analytic $n$ & Early $n$ & Unexposed $n$ & Raw $\Delta$ & IPW $\Delta$ & OW $\Delta$ \\\midrule
Day 14 & 450 & -- & -- & -- & -- & -- \\
Day 30 & 1,820 & 1,624 & 196 & 0.3743 & 0.3861 & 0.3261 \\
Day 45 & 1,726 & 1,599 & 127 & 0.4130 & 0.4738 & 0.3606 \\\bottomrule
\end{tabular}\end{table}

The day-14 analysis removed 1,465 complete cases under the categorical-support restriction, leaving 450 learners, but had fewer than 30 observations in at least one exposure arm; therefore no valid weighted comparison is reported. At day 45, the maximum IPW weight increased to 64.18 and the total IPW effective sample size fell to 314.31 (from 509.52 at day 30), with only 127 unexposed learners. This makes the larger day-45 contrast especially vulnerable to positivity problems and unstable weighting; no dose-response or strengthening causal effect may be inferred. The independent day-30 audit was extended to alternate presentations and post-landmark submission descriptions. Their outputs are reported in the following separate, exploratory section; none supplies causal identification.

\section{Other Course Presentations}
A second successful isolated GitHub Actions execution downloaded the same original UCI archive and verified its checksum, reproducing the focal CCC/2014J estimate again and calculating previously unreported diagnostics and exploratory comparisons (\href{https://github.com/devissaputra/causal_learning_analytics/actions/runs/39686498958}{run 39686498958}). The standalone audit source is available in the \href{https://github.com/devissaputra/causal_learning_analytics/pull/1}{draft research pull request}; the two JSON outputs are downloadable from the run artifact. The audited sample is selected without consulting outcome labels. The secondary comparisons were added during manuscript revision; therefore they should not be interpreted as confirmatory prespecified replications.

\subsection{Arm-wise weight stability and balance}
The day-30 primary IPW weights have effective sample sizes of 1,603.91 among 1,624 early submitters and only 133.81 among 196 non-early submitters. The five largest weights contribute 8.84\% of the total non-early arm weight, and its maximum is 38.64. The alternative overlap weights give arm-wise effective sample sizes 1,045.10 and 192.32. A separately implemented standardized mean-difference diagnostic evaluates 37 numeric/categorical indicators using their weighted within-group variances: maximum absolute SMD falls from 0.5064 unweighted to 0.1161 under IPW (two indicators exceeding 0.10), and 0.0181 with overlap weighting (none exceeding 0.10). These numbers reinforce that overlap weighting balances the \emph{recorded} variables better, but its target population differs and residual or unmeasured confounding remains plausible.

\subsection{Propensity model specifications}
Holding the focal sample and definition fixed, varying logistic regularization to $C=0.25$ gives an IPW contrast of 0.3849; $C=4$ gives 0.3865; and dropping region gives 0.3912. The numerical range is narrow for these particular specifications. It is not a formal sensitivity test of unmeasured confounding or a guarantee against misspecification. In particular, propensity-score extremes and small comparison arms persist.

\subsection{Cross-presentation descriptive replication}
We identified the 17 additional module-presentations with at least 30 learners in each exposure group after the same day-30 complete-case and categorical-support restrictions. We selected the three largest eligible presentations \emph{only by resulting cohort size}, without looking at outcome rates or estimated contrasts. Table~\ref{tab:otherpresentations} reports the resulting point estimates. Each comparison refits its own propensity model, uses a distinct exposure-opportunity context, and is explicitly exploratory; we do not pool effects or claim a universal response to a policy.
\begin{table}[ht]\centering\small
\caption{Additional independently executed day-30 observational contrasts (exploratory). The post-hoc within-presentation bootstrap intervals appear in Table~\ref{tab:crossci}. IPW balance failures remain in all three.}
\label{tab:otherpresentations}
\begin{tabular}{lrrrrrr}\toprule
Presentation & $n$ & No early $n$ & Raw $\Delta$ & IPW $\Delta$ & OW $\Delta$ & IPW ESS\\\midrule
FFF / 2013J & 1,867 & 178 & 0.4933 & 0.5032 & 0.4215 & 445.13\\
BBB / 2013J & 1,804 & 217 & 0.4984 & 0.4768 & 0.4301 & 653.69\\
BBB / 2014J & 1,752 & 96 & 0.5770 & 0.5610 & 0.5416 & 217.39\\\bottomrule
\end{tabular}
\end{table}
\subsection{Full-refit uncertainty for three additional presentations}
After obtaining and inspecting the exploratory point contrasts, we executed a separate, reproducible follow-up that refits the propensity model within each of 1,000 student-level bootstrap samples for each of the same three presentations (UCI raw archive checksum unchanged). The presentation list and cohort definitions were held fixed; this procedure is \emph{post-hoc descriptive uncertainty estimation}, not new prospective confirmation. All 3,000 bootstrap fits succeeded and the numerical point estimates match the independently archived exploratory analysis.
\begin{table}[ht]\centering\small
\caption{Post-hoc 95\% full-refit bootstrap percentile intervals for three further day-30 OULAD presentations. Both contrasts and intervals describe observed submission groups only; IPW and overlap target different weighted populations.}
\label{tab:crossci}
\begin{tabular}{lrrrr}\toprule
Presentation & IPW $\Delta$ & IPW 95\% interval & Overlap $\Delta$ & Overlap 95\% interval\\\midrule
FFF / 2013J & 0.5032 & [0.4479, 0.5476] & 0.4215 & [0.3701, 0.4762]\\
BBB / 2013J & 0.4768 & [0.4126, 0.5368] & 0.4301 & [0.3705, 0.4859]\\
BBB / 2014J & 0.5610 & [0.4636, 0.6324] & 0.5416 & [0.4664, 0.6051]\\\bottomrule
\end{tabular}\end{table}
All intervals are strictly above zero, but they do not account for unmeasured confounding, retention selection, or the exploratory presentation-selection and analysis decisions. These nominal intervals are not adjusted for selection or multiplicity. Especially for BBB/2014J, the exposed/unexposed split is 1,656/96, the maximum IPW weight is 113.67, the IPW effective sample size is 217.39, and the maximum post-IPW absolute SMD is 0.2162. These diagnostics materially limit adjusted comparability despite the nonzero bootstrap interval. Source code, all aggregate estimates and the successful independent workflow are in \url{https://github.com/devissaputra/causal_learning_analytics/actions/runs/39777348662} and \url{https://github.com/devissaputra/causal_learning_analytics/blob/research/paper2-cross-presentation-final-20261022/results/paper2_cross_presentation_bootstrap_20261022.json}.

The primary concern is not whether all contrasts are positive: the comparison sizes, maximum weights, and measured imbalance vary across presentations. Maximum absolute post-IPW SMDs are 0.1135, 0.1236, and 0.2162, respectively; maximum post-overlap SMDs are 0.0456, 0.0350, and 0.0543, using the same within-group-variance convention. The highest IPW contrasts occur under different selection structures, so they should not be interpreted as external causal validation.

\subsection{Why the unexposed category is heterogeneous}
In CCC/2014J, 98 of the 196 learners without an early qualifying submission had a later qualifying submission recorded, and 98 had no qualifying later submission recorded. The final favorable-outcome counts are 32 of 98 (32.7\%) and 0 of 98 (0\%), respectively. These are purely \emph{retrospective} descriptions of groups classified by behavior after day 30. In a course where submitting assessments is closely tied to progression or grading, the absence of later recorded submissions is also structurally entangled with the final outcome. It would be incorrect to use later submission as a baseline confounder, to call the later/never contrast an intervention effect, or to claim that the association is independent of the assessment system's progression requirements.

\section{Identification Boundary and a Testable Future Intervention}
The study does not identify the effect of \emph{making} a learner submit an assessment by day 30. In particular, an observed submission can be both a marker of underlying capacity and part of the grading and progression mechanism. No measured covariate model repairs this structural difference between a behavior and an intervention. The source dataset does not record an independently assigned encouragement, a policy instrument, individualized assessment access or reminder assignment, pre-course learning motivation, or the full chronology of assessment opportunities. Moreover, a hypothetical intervention initiated on day 30 cannot change whether submission occurred in the preceding 30 days; policy assignment and eligibility would have to precede that exposure window.

A prospective follow-up study could instead randomize a \emph{defined, feasible support policy} at or before the beginning of an assessment window, such as offering a structured reminder and optional submission assistance at specified days when an assessment is demonstrably available. Eligibility, assignment, the intention-to-treat target population, outcome horizon, subgroup protections, implementation adherence, and missingness handling would need specification before assignment. Follow-up should retain participants assigned to each policy regardless of subsequent submission, and distinguish effects on submission rates from effects on final course outcomes. Such a design would estimate an effect of \emph{support offered}, not necessarily the effect of compulsory submission. The present OULAD analysis contains no such assignment and cannot be retroactively upgraded to that study.

\section{Discussion}
The adjusted positive association is compatible with several explanations: students with greater preparation or available time may submit earlier and finish successfully; access barriers or employment demands may delay submission and independently worsen final outcomes; assessment schedules and course supports may affect both. Measured background adjustment cannot remove unobserved motivation, prior achievement, workload, digital access, or individualized support. Moreover, retained enrolment to day 30 may itself reflect these factors. The event ``at least one submission'' also conflates timing, opportunity, task type, and degree of completion.

The results therefore support research on monitoring and clarification, not automatic learner risk labeling. OULAD early-warning studies examine calibration, subgroup auditing and cross-presentation transport \citep{wu2026,wufourweeks2026}; these predictive properties still would not identify the benefit of the resulting interventions. Using submission timing as a punitive intervention trigger could disadvantage learners facing structural constraints. Aggregate descriptive analytics with human review and locally evaluated interventions would be more proportionate than policy claims based solely on these historical records.

The missingness/support exclusions depend on measured record properties, not on a missing-at-random guarantee. A comparison of crude contrasts across excluded and retained samples is descriptive; its apparent numerical stability cannot substitute for identification assumptions. The analysis also conditions on being registered and retained to day 30, a potential collider affected by learner circumstances and course supports. Conditional on complete data, all reported point estimates still refer to selected cohorts and cannot be equated with the total enrollment population. This study uses de-identified historical public records; that fact does not eliminate re-identification risk in downstream joined datasets, nor does it license redistribution beyond the dataset terms. The present repository records no human-subject contact or new randomization. Ethical review requirements for later deployed interventions are separate from the reproducibility of this secondary analysis.

\section{Conclusion}
Among CCC/2014J students still enrolled on day 30, those with a qualifying early assessment submission had a substantially higher rate of passing or earning distinction. The positive difference persisted under inverse-probability and overlap weighting, but those estimates describe different weighted populations. Sparse comparison groups, large weights, selection into the day-30 cohort, and unrecorded differences in opportunity or preparation limit what can be inferred. The three additional course presentations show similar descriptive patterns without resolving these problems. These results are useful as an audit of historical learning-analytics associations; they cannot justify an assessment-deadline intervention without a separately designed study.

\section*{Data, code, and verification boundary}
The source is OULAD, UCI dataset 349 (\href{https://doi.org/10.24432/C5KK69}{DOI:10.24432/C5KK69}). The \href{https://github.com/devissaputra/causal_learning_analytics}{public repository} documents the analysis scripts, source provenance, statistical summaries, software dependencies, and independent cohort-reconstruction workflows. The primary 1,820-learner calculation and refitted uncertainty intervals were independently rerun against the checksum-verified UCI source in \href{https://github.com/devissaputra/causal_learning_analytics/actions/runs/39684326436}{Actions run 39684326436}. Additional presentation-level calculations, including refitted bootstrap intervals, are documented in \href{https://github.com/devissaputra/causal_learning_analytics/actions/runs/39777348662}{run 39777348662}. A preserved research snapshot is available at \href{https://github.com/devissaputra/causal_learning_analytics/tree/828e71a7420519ac7c5c3306669c9402ca8f4064}{commit 828e71a}. These analyses verify specified numerical calculations in historical records; they do not establish unconfounded causal effects. No individual student records are redistributed. The present semantic \LaTeX{} manuscript and its figures form a separate 7 November 2026 editorial source edition.

\section*{Generative AI assistance and author accountability}
OpenAI's ChatGPT was used substantially in developing this work, including preparation and revision of manuscript drafts, assistance with analysis code, explanatory presentation of methodological decisions, scientific-document formatting, and technical checks. The empirical findings are anchored to documented scripts, public data provenance, and recorded independent calculations; language-model output alone is not evidence of a numerical result. Generative AI is not an author. Devis Wawan Saputra, the named human author, approved the stated scientific interpretations and accepts responsibility for the methods, accuracy, originality, sources, and publication disclosures. The analysis has not established a causal policy effect and has not undergone independent peer review.

\end{document}